# Optical second harmonic generation in a ferromagnetic liquid crystal

Jure Brence,[a] Luka Cmok,[b] Nerea Sebastián,[b] Alenka Mertelj,[b] Darja Lisjak,[c] Irena Drevensek-Olenik[*ab]

[a] *University of Ljubljana, Faculty of Mathematics and Physics, Jadranska 19, SI-1000, Ljubljana, Slovenia. E-mail: irena.drevensek@ijs.si*
[b] *Jožef Stefan Institute, Department of Complex Matter, Jamova 39, SI-1000 Ljubljana, Slovenia.*
[c] *Jožef Stefan Institute, Department for Materials Synthesis, Jamova 39, SI-1000 Ljubljana, Slovenia.*

A comparative experimental investigation of the dependence of second harmonic generation (SHG) on applied external voltage between the standard nematic liquid crystalline material and an analogue ferromagnetic nematic liquid crystalline material was performed by using a fundamental optical beam at 800 nm wavelength. For a ferromagnetic material, the dependence of SHG on an applied magnetic field was also examined. Three different polarization combinations of the fundamental and the second harmonic radiation were analysed. The SHG signal observed in the former material is attributed to a combination of electric field-induced SHG (EFISHG) and flexoelectric deformation-induced SHG, while SHG signal observed in the latter material is attributed solely to flexoelectric deformation-induced SHG. The obtained dependences of the SHG signal on the associated optical retardation show that in the most favourable polarization combination the two contributions generate about the same effective nonlinear optical susceptibility.

## Introduction

Optical second harmonic generation (SHG) is an important experimental method for resolving structural polarity of different materials. In this work we present an example of the SHG study performed on a soft matter system. We investigate nematic liquid crystals (LCs) that are centrosymmetric materials and consequently the process of optical second harmonic generation (SHG) in electric dipole approximation is forbidden in a uniform nematic phase.[1] However, because typical LC molecules possess a non-centrosymmetric molecular structure,[2] any mechanism that breaks the centrosymmetry of the LC medium enables also the appearance of the SHG. One of such mechanisms is electric poling, which is based on the property that application of an external DC or low frequency electric field on the centrosymmetric medium causes an asymmetric (polarized) electronic response to the optical field.[3] The corresponding SHG process, known as electric field-induced second harmonic generation (EFISHG), was first observed in LCs around 1980.[4-8] Another mechanism that can break the centrosymmetry is the flexoelectric effect, in which spatially inhomogeneous orientational arrangement of the LC molecules induces local polarization of the LC medium.[9,10] This source of SHG in LCs was discovered at about the same time as EFISHG.[11,12] The flexoelectric effect is responsible for occurrence of SHG in the vicinity of topological defects and other imperfections in the LC structure.[13, 14] It also causes the appearance of scattered second harmonic (SH) radiation provoked by thermally-driven orientational fluctuations of the LC molecules (SH scattering).[15,16] Another source of SHG are contact regions of the LC medium with surrounding surfaces, where censtrosymmetry is intrinsically absent. This property, which is unavoidable in practical situations, leads to the so-called surface SHG (SSHG).[17] Experiments with various LC systems demonstrated that SSHG can be used as a very convenient tool for studying LC surface alignment mechanisms.[18-20] Besides the above described sources, there exist quadrupolar and other multipolar mechanisms that can cause SHG even in the uniform nematic phase. But, their magnitude is in general much lower.[21,22]

When a LC layer is exposed to an external electric field oriented in direction perpendicular or nearly perpendicular to the LC alignment direction, which is a common configuration in various LC-based technological devices, the applied field causes an inhomogeneous reorientation of the LC molecules and consequently enables both, EFISHG and flexoelectric deformation-induced SHG. Different research groups tried to separate the two contributions from each other. Their approaches included time resolved SHG measurements after switching on and off of the applied voltage,[23,24] application of short voltage pulses,[25,26] and usage of specially designed waveforms with zero-voltage intervals.[27] It was found that EFISHG was practically a momentary effect, while flexoelectric deformation-induced SHG followed the kinetics of the LC reorientation process. Another observation was that the two SHG contributions emerged in different polarization combinations of the fundamental and the SH field. Therefore they can in principle be separated by a suitable choice of polarizations. However, this separation effect is at present not yet fully understood. Besides this, there still exist also some other important open questions, in particular the problem of constructive interference of the SH radiation, i.e. the possibility to realize the so-called phase matching condition.[1,26]

This work investigates the SHG process in a ferromagnetic nematic LC phase. Such a phase can be realized by dispersion of platelet-shaped ferromagnetic nanoparticles in a conventional nematic LC material.[28,29] One of its most profound characteristics is that its orientational structure can be manipulated by magnetic fields as low as a few mT. A comparative analysis of the SHG signal as a function of applied voltage in standard LC material (E7) and in the same LC material with addition of nanoplatelets is performed. Furthermore, the dependence of the SHG signal on the applied magnetic field is examined for the ferromagnetic phase. As magnetization induced SHG (MSHG) is believed to be negligible,[30-32] the SHG observed in the latter case is attributed to magnetic field-driven flexoelectric deformation. A basic theoretical framework for description of the corresponding SHG processes that takes into account spatially varying nonlinear optical susceptibility associated with spatially varying orientational profile of the LC material is also presented.

## Materials and methods

A ferromagnetic nematic liquid crystalline material was composed of the commercial LC mixture E7 (Merck Ltd.) and of ferromagnetic scandium doped barium hexaferrite ($BaFe_{12-x}Sc_xO_{19}$, BaHF) single-crystal nanoplatelets. The details on material preparation are described elsewhere.[28,29,33] A selected LC compound (either standard E7 or its ferromagnetic variant) was introduced via capillary action into commercial LC cells (Instec, Inc.) composed of ITO-coated glass plates covered with a rubbed polyimide layer. The cells are designed to generate planar LC alignment with a small pretilt angle of a few degrees.[2] The thickness of the LC layer was $D$=20.4 μm. In case of ferromagnetic LCs, the samples initially exhibited a polydomain magnetic structure. So, a magnetic field $B \leq 50$ mT oriented along the rubbing direction was used to transform them to monodomain ones. The uniformity of the resulting structure was inspected by transmission polarization optical microscopy (POM). All SHG and other measurements were performed at room temperature.

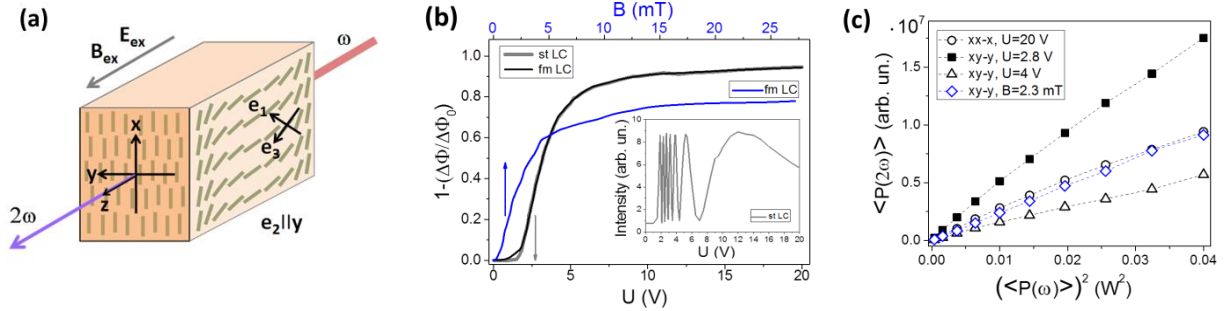

Fig. 1 (a) Schematic drawing of nematic director field in the investigate LC cells. Rubbing direction is parallel to the *x*-axis. Fundamental and SH optical beams are propagating along the *z*-axis. External electric or magnetic field are applied along the *z*-axis. Field-induced reorientation of LC molecules ocurrs in the *xz* plane. The local axis of symmetry **e₃** is parallel to the nematic director **n**. The local axis **e₁** is perpendicular to **n** and lies in the *xz* plane.The local axis **e₂** is parallel to the *y*-axis. (b) Relative variation of optical retardation  1-($\Delta\Phi/\Delta\Phi_0$) as a function of appllied voltage and as a function of applied magnetic field. $\Delta\Phi_0$ is the initial retardation in zero field. Grey line corresponds to standard nematic phase (denoted as st) and black and blue lines to the ferromagnetic (denoted as fm) nematic phase. The inset shows an example of the transmitted intensity as a function of applied voltage.  (c) Measured average SH power <$P(2\omega)$> from the ferromagnetic nematic sample as a function of the square of the average incident optical power <$P(\omega)$>. Different symbols correspond to measurements performed with different polarization combinations and with different fields applied to the sample, as donoted in the figure.

Fig. 1a shows a schematic drawing of the LC layer and the designation of the coordinate system. The rubbing direction corresponds to the *x*-axis and glass plates are parallel to the *xy* plane. Local orientation of LC molecules in the nematic phase is described by the unit vector **n** called the director. Since directions **n** and −**n** have the same probability, molecular orientational distribution is described by the tensor order parameter **Q**=$S$(**n**⊗**n**-**I**/3), where $S$ is the scalar order parameter dependent on the degree of orientational ordering along **n**.[2] The nematic phase exhibits symmetry characteristics of the point group D$_{\infty h}$, among them also the inversion symmetry. From the perspective of optical properties, nematic LCs are uniaxial optically birefringent materials with optical axis parallel to **n**. When an external electric or magnetic field oriented in direction perpendicular to the glass plates (*z*-axis in Fig. 1a) is applied to the LC cell, the director and consequently also the optical axis tend to orient along the direction of the field. This effect is opposed by boundary conditions at the glass surfaces that try to preserve the initial orientation. As a result of this competition, an inhomogeneous orientational profile is established that can be described as **n**=(*cos*($\theta(z)$),0, *sin*($\theta(z)$)), where $\theta(z)$ is the inclination angle with respect to the alignment axis *x*. The reorientation process leads to modification of the optical retardation between the extraordinary and the ordinary optical beams

$$\Delta\Phi = \left(\frac{\omega}{c_0}\right) \int_0^D [n_e(\theta(z)) - n_o] \, dz \, , \tag{1}$$

where $\omega$ is angular frequency of the optical radiation, $c_0$ is the light speed in vacuum, $n_e$ is the refractive index of the extraordinary beam that depends on $\theta(z)$ as

$$\left(\frac{1}{n_e}\right)^2 = \left(\frac{\cos(\theta(z))}{n_{||}}\right)^2 + \left(\frac{\sin(\theta(z))}{n_{\perp}}\right)^2 \qquad (2)$$

and $n_o = n_{\perp}$ is the refractive index of the ordinary beam.[1] Here $n_{||}$ and $n_{\perp}$ are principal refractive indices of the LC medium. Measurements of $\Delta\Phi$ as a function of applied field are typically used to obtain a characteristic fingerprint of the field-induced LC reorientation processes.

Optical retardation as a function of applied voltage $U$ (1 kHz square wave form) or magnetic field $B$ (DC electromagnet) was resolved by the conventional method based on measurements of sample transmittance between two crossed polarizers oriented at 45° with respect to the rubbing direction.[34] A low power He-Ne laser with a wavelength of 633 nm was used as a light source. The obtained results are shown in Fig. 1b, where $\Delta\Phi_0$ corresponds to the initial retardation in zero field. The inset shows an example of the measured transmitted intensity as a function of applied voltage.

Optical SHG measurements were performed with a pulsed Ti:sapphire laser operating at a wavelength of 800 nm (Legend Elite, Coherent). The pulses were ~100 fs long and had a repetition rate of 1 kHz. The spot size of the laser beam on the sample was 2 mm. The beam was linearly polarized either along the *x*-axis, at 45° with respect to the *x*-axis, or along the *y*-axis (see Fig. 1a). A low pass optical filter was placed in front of the sample to block any SH radiation that was generated before the beam entered the sample. A high pass filter was placed behind the sample to block the fundamental radiation and consequently to prevent any additional SHG on optical components positioned behind the sample. The SH beam generated in transmission direction was sent onto the analyser oriented either along the *x* or along the *y*-axis. The corresponding polarization combinations of the fundamental ($\omega$) and the SH radiation ($2\omega$) are denoted as *ij-k*, where the first two indices indicate the fundamental and the last index the SH field. For instance, the combination *xy-y* describes a case in which the fundamental beam is polarized at 45° with respect to the *x*-axis, so that it involves the *x*- and the *y*-polarized fundamental field, and the SH beam is polarized along the *y* axis.

The investigated SH radiation was resolved from the background of SH scattering and other stray light by a grating spectrograph (Acton Spectra Pro, Priceton Instruments). The spectrum of outgoing radiation was detected with a gated image intensified CCD camera (iStar, Andor Technology). The gating was synchronized with the laser pulses. After background subtraction, the average generated SH power $<P(2\omega)>$ was obtained by integration of the signal detected in 20 nm broad interval around the SH wavelength. The setup was tested by measuring a dependence of $<P(2\omega)>$ on average incident optical power $<P(\omega)>$ in a ferromagnetic LC sample exposed to different applied external fields. The results are shown in Fig. 1c. For all tested cases, a proper quadratic dependence was observed up to $<P(\omega)>$=200 mW, indicating that no optical field-induced sample reorientation or bleaching occurred. Therefore, the rest of the measurements were performed at $<P(\omega)>$=200 mW. With this incident power a typical accumulation time for the SHG signal was 40 s.

Because the investigated samples contained ITO electrodes that are known to produce relatively strong SHG,[35] all measurements were performed at normal incidence of the fundamental beam with respect to the ITO surfaces, i.e. the incident and the SH beam propagated along the *z*-axis in Fig. 1a. In this case, due to symmetry considerations, the SHG signal from ITO films should be zero, which was verified by measurements performed on empty cells. When an empty cell was rotated around the *y* axis, SHG signal appeared and <*P*(2*ω*)> obtained for incident and SH beams linearly polarized in the *xz* plane (also known as *p*-polarization) monotonously increased with the increasing rotation angle. We decided to use the value of <*P*(2*ω*)> measured on the empty cell at the incident angle of 45° as a reference, i.e. all values for <*P*(2*ω*)> reported for LC-filled cells are given relatively with respect to that reference.

## Results and discussion

SHG measurements

Fig. 2 shows the results of SHG measurements obtained in three polarization combinations, in which the coherent SHG signal could have been clearly resolved from the background. These are, listed with respect to the decreasing signal magnitude, the combinations *xy-y*, *xx-x* and *yy-x*. Black symbols display dependencies of <*P*(2*ω*)> on applied voltage *U*. Open circles are data for the standard nematic material (labelled as st) and full circles for the ferromagnetic material (labelled as fm). In all polarization combinations both materials exhibit non monotonic behaviours. Blue symbols (diamonds) display dependencies of <*P*(2*ω*)> on the applied magnetic field *B*. Only results for the ferromagnetic material are given, as the standard nematic material was completely nonresponsive to the low *B* fields used in the experiments. Significant changes of <*P*(2*ω*)> as a function of *B* can be noticed only in the *xy-y* polarization combination, while in the other two polarization combinations the observed modifications are quite minor.

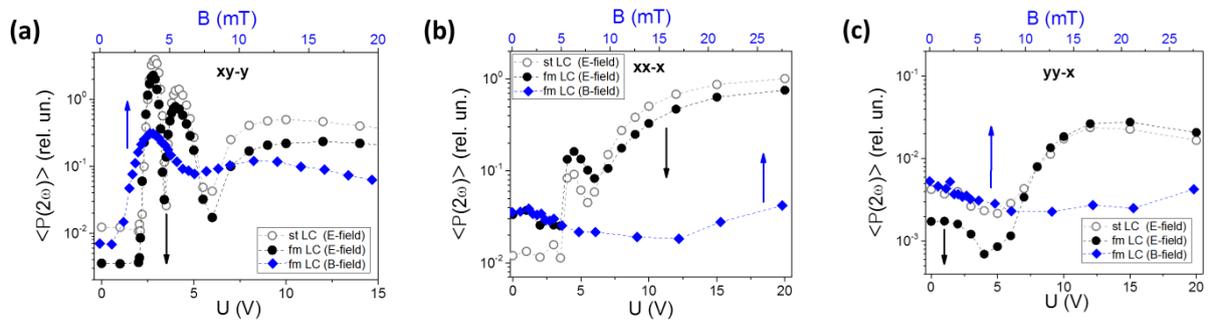

Fig. 2 SHG signal as a function of applied voltage *U* (black circles) and as a function of applied magnetic field *B* (blue diamonds) for three different polarization combinations of the fundamental and the SH light: (a) *xy-y*, (b) *xx-x*, (c) *yy-x*. Open circles are data for standard nematic material (pure E7) exposed to applied voltage *U*. Solid circles are data for ferromagnetic nematic material (E7 with addition of ferromagnetic nanoplatelets) exposed to applied voltage *U*. Blue diamonds are data for ferromagnetic nematic material exposed to applied magnetic field *B*. Dashed lines are guides to the eye.

It is expected that field-induced modifications of <*P*(2*ω*)> are correlated with field-induced reorientation of the LC. As mentioned before, the latter can be resolved from the observed field-induced modifications of optical retardation $\Delta\Phi$ (Fig. 1b). Therefore, for a more appropriate comparative analysis it is better to present the results for <*P*(2*ω*)> shown in Fig. 2 as a function of relative variation of optical retardation $1-(\Delta\Phi/\Delta\Phi_0)$. In addition, for easier notion of the main differences between different cases it is more convenient to plot <P(2ω)> in linear instead of logaritimic scale. With this in mind, the obtained data were grouped into two separate sets, the first one showing the SHG signal as a function of voltage-induced reorientation for the standard and the ferromagnetic nematic material and the second one showing the SHG signal from the ferromagnetic nematic material as function of voltage-induced and as a function of magnetic field-induced retardation modification. The two sets are shown in Figs. 3 and 4.

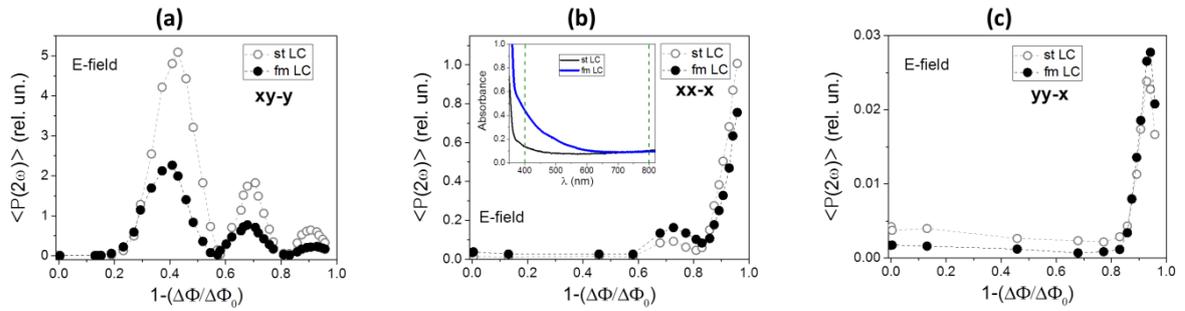

Fig. 3 SHG signal as a function of relative variation of voltage-induced optical retardation $1-(\Delta\Phi/\Delta\Phi_0)$ for standard nematic material (open circles) and for ferromagnetic nematic material (solid circls). The results obtained in different polarization combinations are shown: (a) *xy-y*, (b) *xx-x* and (c) *yy-x*. The inset in (b) shows optical absorption spectra of both materials. The thinner black line corresponds to the standard and the thicker blue line to the ferromagnetic nematic material.

As can be noticed in Fig. 3, the dependencies of the SHG signal on voltage-induced retardation modification for the standard and the ferromagnetic nematic material are very similar in all polarization combinations. Just the magnitude of the signal is sometimes larger for the former and sometimes for the latter. The ferromagnetic phase contains a small concentration of ferromagnetic nanoparticles (< 0.1 vol %),[36] which increases optical absorption in the short-wavelength region of the spectrum. Absorption spectra of the two analogous samples, one filled with the standard and one with the ferromagnetic material, are shown in the inset of Fig. 3b. At the fundamental wavelength (800 nm) both materials are similarly transparent, while at SH wavelength (400 nm) the ferromagnetic material is less transparent than the standard one. Consequently, one would expect that <P(2ω)> from the ferromagnetic material is notably smaller than from the standard one, which is indeed the case in the *xy-y* polarization combination. The maximal values of <P(2ω)> for both materials are observed at $(1-(\Delta\Phi/\Delta\Phi_0))\sim0.4$. For the other two polarization combinations maximal SHG signals appear at final stages of the reorientation process, i.e. at $(1-(\Delta\Phi/\Delta\Phi_0))\sim1$, and differences are much smaller. Evident similarity between the SHG responses of both materials signifies that addition of magnetic nanoplatelets to the standard LC material practically does not affect its nonlinear optical response. This is in agreement with the results on optical retardation $\Delta\Phi$ (*U*) (Fig. 1b), which are also very similar for both materials.

In contrast, as can be seen in Fig. 1b, the variations $\Delta\Phi$ (*U*) and $\Delta\Phi$ (*B*) for the ferromagnetic nematic material are very different. In the latter case there is no threshold present, but a profound molecular reorientation starts already at very low values of *B*. Besides this, in large *B* fields the reorientation saturates at $(1-(\Delta\Phi/\Delta\Phi_0))\sim0.8$, instead at $(1-(\Delta\Phi/\Delta\Phi_0))\sim1.0$. Those differences were discussed already in our previous work,[36] so here the emphasis is put on the differences associated with the SHG signal. Those can be resolved in Fig. 4. One can notice that in the *xy-y* polarization combination the dependencies of <P(2ω)> on voltage- and on magnetic field-induced retardation modification ΔΦ are very similar, only the magnitude of the signal obtained with applied magnetic field is about four times smaller from the one obtained with applied voltage. On the contrary, for the other two polarization combinations the SHG signal observed in *B* field is quite small and practically constant at all accessible values of 1-(ΔΦ/ΔΦ$_0$).

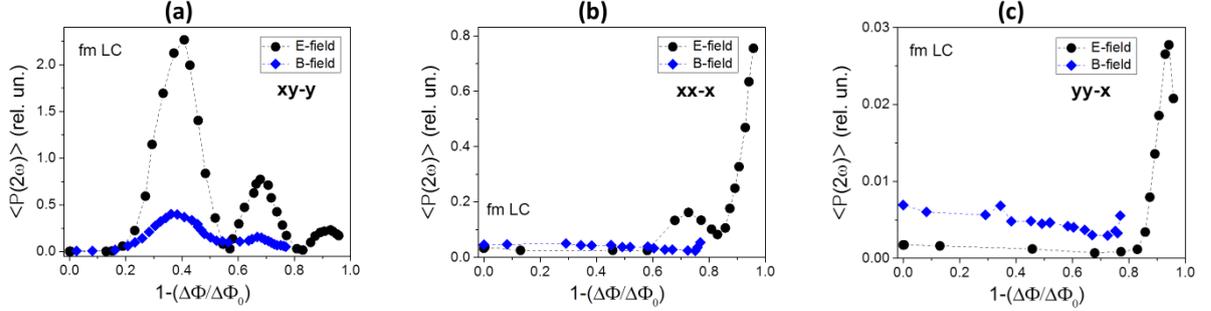

Fig. 4 SHG signal from ferromagnetic nematic phase as a function of voltage-induced (black circles) and as a function of magnetic field-induced (blue diamonds) relative variation of optical retardation 1-(ΔΦ/ΔΦ$_0$). The results obtained in diferent polarization combinations are shown: (a) *xy-y*, (b) *xx-x* and (c) *yy-x*.

Theoretical analysis

In the absence of any applied external voltage or magnetic field the observed SHG signal from the investigated samples was practically indistinguishable from the background associated with the SH scattering. Therefore we conclude that quadrupolar and other higher order SHG contributions are negligible in our experimental configuration. So, the dominating SHG source is assumed to be the nonlinear induced dielectric polarization[1]

$$\mathbf{P}_{NL}(2\omega) = \varepsilon_0 \chi^{(2)} : \mathbf{E}^*(\omega)\mathbf{E}^*(\omega) \ , \tag{3}$$

where $\varepsilon_0$ is vacuum permittivity, $\chi^{(2)}$ is the tensor of the second-order nonlinear optical susceptibility and $\mathbf{E}^*(\omega)$ is the local optical field at fundamental frequency $\omega$. The main origins of $\chi^{(2)}$ are applied external voltage and flexoelectric deformation. Both of them break the equivalence between the $\mathbf{P}_{NL}$ induced along the +**n**(**r**) and –**n**(**r**) directions, which corresponds to a reduction of the local symmetry from the point group $D_{\infty h}$ to $C_{\infty v}$. Consequently, the LC material remains optically uniaxial with an optical axis parallel to **n**(**r**), but it is not anymore centrosymmetric. As a result of this, in the local coordinate system $\mathbf{e}_1$, $\mathbf{e}_2$, $\mathbf{e}_3$ in which $\mathbf{e}_3$ is parallel to **n**(**r**) (see Fig. 1a), there emerge the following nonzero components of $\chi^{(2)}$: $\chi_{333}$; $\chi_{311}=\chi_{322}$ and $\chi_{131}=\chi_{232}$.[37] Further, by considering the Kleinmann symmetry,[38] it follows that $\chi_{311}=\chi_{322}=\chi_{131}=\chi_{232}$, so there exist only two independent components, namely $\chi_{333}$ and $\chi_{311}$.

Both, $\chi_{333}$ and $\chi_{311}$, are supposed to be proportional to the effective local poling field $\mathbf{E}_p(\mathbf{r})$,[37] which is determined by the projections of the low-frequency electric field and of the flexoelectric polarization onto the nematic director **n**(**r**):

$$\mathbf{E}_p = C_E\big(\mathbf{E}^*_{lf} \cdot \mathbf{n}\big)\mathbf{n} + C_F(\nabla \cdot \mathbf{n})\mathbf{n} = \left[C_E E^*_{lf}\sin(\theta(z)) + C_F\left(\frac{d\theta(z)}{dz}\right)\cos(\theta(z))\right]\mathbf{n} \ , \tag{4}$$

where $\mathbf{E}_{lf}^*$ is the local low frequency electric field generated by an external voltage $U$ applied along the z-axis (Fig. 1a), $(\nabla \cdot \mathbf{n})\mathbf{n}$ is the flexoelectric poling field associated with the splay-type distortion of $\mathbf{n(r)}$,[9,10,14] and $C_E$ and $C_F$ are constants. In a distorted LC structure, such as the one depicted in Fig. 1a, the first term in Eq. 4 is maximal in the central region of the LC cell, while the second term is maximal in the vicinity of the glass substrates. If, instead of a voltage, a magnetic field is used to induce the reorientation, only the second term remains.

An additional simplification arises, if we assume that molecular hyperpolarizability of the LC molecules is strongly predominated by the component $\beta_{\xi\xi\xi}$ parallel to the long molecular axis $\mathbf{e}_\xi$, which is a reasonable approximation for biphenyl-type LC molecules, such as those forming the E7 mixture used in our work.[39] In this approximation it follows:[37]

$$\chi_{333} \propto \left(\frac{1}{5} + \frac{4}{7}\langle P_2 \rangle + \frac{8}{35}\langle P_4 \rangle\right) E_p \beta_{\xi\xi\xi} \; ; \quad \chi_{311} \propto \left(\frac{1}{15} + \frac{1}{21}\langle P_2 \rangle - \frac{8}{70}\langle P_4 \rangle\right) E_p \beta_{\xi\xi\xi} \quad , \tag{5}$$

where $\langle P_2 \rangle$ and $\langle P_4 \rangle$ are orientational order parameters associated with axially symmetric orientational distribution of LC molecules around $\mathbf{n}$, which are assumed to remain constant during the reorientation process.[2] From Eq. 5 it follows that $\chi_{333}$ is in general significantly larger than $\chi_{311}$.

To obtain the effective susceptibilities relevant in the utilized experimental configuration (Fig. 1a), one needs to transform $\chi^{(2)}$ from the local coordinate system 1, 2, 3 to the laboratory system $x, y, z$. After this transformation the effective susceptibilities for the investigated polarization combinations can be written as

$$\chi_{xy-y} = 2\chi_{yy-x} = 2\chi_{311} \cos\theta(z) \,, \tag{6}$$

and

$$\chi_{xx-x} = \chi_{333} \cos^3\theta(z) + 3\chi_{311} \cos\theta(z) \sin^2\theta(z) \,. \tag{7}$$

In the calculation of the SH optical field $\mathbf{E}^*(2\omega)$, besides $E_p$ and $\chi_{ij-k}$, there arises an additional term dependent on $\theta(z)$, namely a wave vector mismatch between the fundamental and the SH field that is given as

$$\Delta k_{xy-y}(z) = \frac{2\omega}{c_0} n_o(2\omega) - \frac{\omega}{c_0} n_o(\omega) - \frac{\omega}{c_0} n_e(\omega, \theta(z)) \,, \tag{8}$$

$$\Delta k_{yy-x}(z) = \frac{2\omega}{c_0} n_e(2\omega, \theta(z)) - \frac{2\omega}{c_0} n_o(\omega) \,, \tag{9}$$

$$\Delta k_{xx-x}(z) = \frac{2\omega}{c_0} n_e(2\omega, \theta(z)) - \frac{2\omega}{c_0} n_e(\omega, \theta(z)) \,, \tag{10}$$

where color dispersion of refractive indices has to be taken into account. The investigated LC material (E7) possesses a positive optical anisotropy (($n_\parallel - n_\perp$)>0) and a normal color dispersion ($n_e(2\omega) > n_e(\omega)$, $n_o(2\omega) > n_o(\omega)$), from which it follows $\Delta k_{yy-x} > 0$ and $\Delta k_{xx-x} > 0$ for every value of $\theta(z)$, while $\Delta k_{xy-y}$ can become zero at a specific value of $\theta(z)$. In a spatially homogeneous medium, in which $\chi_{ij-k}$ and $\Delta k_{i-jk}$ are constant, the latter situation corresponds to the so-called type-II phase matching situation and is associated with maximal conversion efficiency from fundamental to SH radiation.[1]

In case of space-varying $\chi^{(2)}(\mathbf{r})$ connected with the inhomogeneous LC reorientation processes, the condition for obtaining maximal conversion efficiency is in general much more complex, because not only $\chi^{(2)}$, but also linear optical susceptibility $\chi^{(1)}$ (associated with refractive indices $n_e$ and $n_o$) is space dependent. One possible approach to analyze such situations is to expand the effective susceptibilities relevant for particular polarization combination into the Fourier series with respect to $q=2\pi/D$, which mainly makes sense only for periodically modulated structures. In such cases maximal SH output is obtained when a relation

$$\Delta k_{ij-k} = mq \ , \quad m = 0, \pm 1, \pm 2, \dots , \tag{11}$$

is valid. For $m \neq 0$ this type of phase matching is known as a so-called umklapp phase matching that is observed in helically twisted LC phases, such as the cholesteric (N*) or ferroelectric smectic (SmC*) phase.[40-42] In other situations the nonlinear wave equation for the SH optical field $\mathbf{E}^*(2\omega,\mathbf{r})$ that is given as[1]

$$\nabla^2 \mathbf{E}^*(2\omega,\mathbf{r}) + \left(\frac{2\omega}{c_0}\right)^2 (2\omega)(\mathbf{1}+\chi^{(1)}(\mathbf{r}))\mathbf{E}^*(2\omega,\mathbf{r}) = -\left(\frac{2\omega}{c_0}\right)^2 \chi^{(2)}(\mathbf{r}):\mathbf{E}^*(\omega,\mathbf{r})\mathbf{E}^*(\omega,\mathbf{r}) , \tag{12}$$

has to be solved. In reorientation processes associated with stratified structures, such as the one investigated in our experiments (Fig. 1a), $\chi^{(1)}$ and $\chi^{(2)}$ vary only along one direction. In this case for solving Eq. 12 the standard Berreman 4×4 matrix formalism, which is widely used to analyze linear light propagation through the LC cells, can be extended to generation and propagation of the SH field.[43,44] However, this approach is beyond the scope of this work and will be reported in a separate paper.

To be able to judge on the importance of the above-described formalism with respect to the usual approximation assuming homogeneous reorientation of the LC director field, i.e. reorientation in which the reorientation angle $\theta$ is independent of $z$, quantitative data on optical refractive indices of the investigated LC material have to be considered.[45] By taking into account the data for E7 reported in (ref. 45) ($n_{||}$(400 nm)=1.84, $n_{\perp}$(400 nm)=1.55; $n_{||}$(800 nm)=1.70, $n_{\perp}$(800 nm)=1.50), we calculated that type the II phase matching condition, which is fulfilled when

$$n_o(2\omega) = \frac{1}{2}n_o(\omega) + \frac{1}{2}n_e(\omega,\theta) , \tag{13}$$

should occur at $\theta$=42°, which, at the He-Ne laser wavelength of 633 nm for which $n_{||}$(633 nm)=1.74 and $n_{\perp}$(633 nm)=1.52, [45] corresponds to 1−($\Delta\Phi/\Delta\Phi_0$) = 0.51. As can be seen in Figs 3a and 4a, in the experiments maximum values of <$P(2\omega)$> were obtained at 1−($\Delta\Phi/\Delta\Phi_0$) ∼ 0.4. For other reorientation angles, <$P(2\omega)$> as a function of $\Delta\Phi$ should exhibit the so-called Maker oscillations whose periodicity is given by the relation $\Delta\Phi_M=\Delta k_{xy-y}\cdot D$= 2π.[46] At the He-Ne laser frequeny this gives $\Delta\Phi$ =($\omega_{\text{He-Ne}}/\omega$)·2π that correponds to ($\Delta\Phi/\Delta\Phi_0$)=0.19, which also quite satisfactory matches the experimental observations. So we can conclude that in the investigated system, the approximation with a homogeneous LC reorientation gives a good overall clue on the observed features, while to describe the details one needs to take into account an appropriate spatial variation of the rotation angle $\theta(z)$ and the associated spatial variations of the linear and nonlinear optical susceptibilities.

## Conclusions

Our results demonstrate that, besides stronger absorption of SH radiation, addition of ferromagnetic BaHF nanoplatellets to the standard nematic liquid crystalline material practically does not affect its nonlinear response to fundamental optical radiation at 800 nm wavelength. This signifies very good dispersion of the platelets in the LC medium and also their minor influence on linear as well as nonlinear optical susceptibility of the host material. This finding is opposite to a recently reported SHG study on dispersion of ferroelectric nanoparticles (BaTiO$_3$) in the nematic material (5CB), in which photo-stimulated polarization of the particles was found to play an important role in the SHG process.[47] In our samples, at incident optical intensity used in the experiments (~0.05 W/mm$^2$), no optical-field induced effects were detected. But, we noticed that long exposures of the samples to low frequency external electric field caused redistribution of magnetic particles and subsequent formation of some localized defects, which typically relaxed to the initial state several hours after switching off the voltage. This is to our opinion the main reason why the values of <$P(2\omega)$> detected in zero field are not always exactly the same (Fig. 2).

The superimposing of strain- and electric-field induced SHG processes investigated in our work is not unique for liquid crystals, but is observed also in centrosymmetric crystalline media. Many such investigations are focused on silicon (Si) and related materials, because of their importance in photonics applications.[48-50] Also in crystalline media it is rather difficult to resolve different sources that contribute to the observed SHG signal.[51] In the ferromagnetic nematic LC medium, investigated in our experiments, we tried to distinguish between the EFISHG and other contributions to the SHG signal by inducing LC reorientation one time with external electric field and another time with external magnetic field. In the latter case, the EFISHG contribution is assumed to be absent. According to this, it can be concluded from Fig. 4a that in the *xy-y* polarization combination flexoelectric deformation alone gives only about ¼ of the total SHG signal as obtained when both, EFISHG and flexoelectric deformation-induced SHG, take place. This means that both effects generate about the same contributions to the effective nonlinear optical susceptibility $\chi_{xy-y}$. In the *xx-x* polarization combination, the contribution from the flexoelectric deformation seems to be practically zero. This is in agreement with the findings reported in (ref. 27), where maximal non-EFISHG-based SHG was also observed in the *xy-y* polarization combination. Somewhat unexpected situation occurs in the *yy-x* polarization combination, in which <$P(2\omega)$> detected for applied magnetic field is larger than <$P(2\omega)$> detected for applied electric field. But, the values of <$P(2\omega)$> in this polarization combination are very low and the differences appears already in zero field, so we attribute them to slightly different initial orientational structure of the sample.

Although the difference between magnetic field-induced and electric field-induced SHG signals observed in the *xy-y* polarization combination is prominent, the corresponding SHG contribution attributed to the flexoelectric deformation might still not be fully reliable. Due to different coupling mechanisms between the nematic director field **n**(**r**) and external magnetic and electric fields, field-induced reorientation profiles $\theta(z)$ that give the same optical retardation $\Delta\Phi$ are not necessary exactly the same. As different reorientation profiles result in different flexoelectric deformation, the associated SHG contributions might be different too. Certainly, further investigations of the problem, such as SHG measurements at oblique incident angles of fundamental and SH optical beams, are needed to improve the understanding of the associated phenomena. For this purpose, LC cells without ITO layers should be used to avoid strong SHG contribution from the ITO films. However, in this case comparative studies between magnetic and electric field-induced SHG processes are expected to be more complicated, because nonplanar electrode configurations usually give less homogeneous electric fields. Besides this, there also remain many other interesting open problems to be studied in ferromagnetic nematic phase, for instance, the effect of magnetic domain structure on the SHG signal or the effect of a magnetic field on the SH scattering.

**Acknowledgements**

The authors acknowledge financial support from Slovenian research agency (ARRS) (grants no. P1-0192, P2-0089-4 and J7-8267) and Slovenian Ministry of Education, Science and Sport (MIZŠ) & ERDF (project OPTIGRAD).